# Gate-Tunable Critical Current of the Three-Dimensional Niobium Nano-Bridge Josephson Junction

*Shujie Yu[1,2], Lei Chen[1,2,\*], Yinping Pan[1], Yue Wang[1,2], Denghui Zhang[1,2], Guangting Wu[1,2], Xinxin Fan[1,2], Xiaoyu Liu[1], Ling Wu[1], Lu Zhang[1,2], Wei Peng[1,2], Jie Ren[1,2], and Zhen Wang[1,2,3,\*]*

[1] National Key Laboratory of Materials for Integrated Circuits, Shanghai Institute of Microsystem and Information Technology(SIMIT), Chinese Academy of Sciences, Shanghai 200050, China

[2] University of the Chinese Academy of Sciences, Beijing 100049, China

[3] School of Physical Science and Technology, Shanghai Tech University, Shanghai 200031, China

*Corresponding authors: leichen@mail.sim.ac.cn, zwang@mail.sim.ac.cn

**ABSTRACT**

Recent studies have shown that the critical currents of several metallic superconducting nanowires and Dayem bridges can be locally tuned using a gate voltage ($V_g$). Here, we report a gate-tunable Josephson junction structure constructed from a three-dimensional (3D) niobium nano-bridge junction (NBJ) with a voltage gate on top. Measurements up to 6 K showed that the critical current ($I_c$) of this structure can be tuned to zero by increasing $V_g$. The critical gate voltage $V_{gc}$ was reduced to 16 V and may possibly be reduced further by reducing the thickness of the insulation layer between the gate and the NBJ. Furthermore, the flux modulation generated by Josephson interference of two parallel 3D NBJs can also be tuned using $V_g$ in a similar manner. Therefore, we believe that this gate-tunable Josephson junction structure is promising for superconducting circuit fabrication at high integration levels.

Interest in superconducting circuits has been rising because of their combination of ultra-high speed with ultra-low power consumption, particularly when conventional complementary-metal-oxide-semiconductor (CMOS) technology has been reaching its physical limits.[1, 2] Although the fabrication of large-scale superconducting circuits based on Nb/AlO$_x$/Nb junction technology is progressing rapidly[3-5], the circuit integration level remains much lower than that of CMOS technology. Unlike field-effect transistors (FETs) in CMOS circuits, conventional superconducting Josephson junctions (JJs) cannot be controlled using a local gate. In circuit applications, the superconducting quantum interference device (SQUID), which consists of two parallel JJs, is usually treated as an equivalent JJ with a tunable critical current $I_c$[6], because an on-chip field coil can be integrated to modulate the $I_c$ of a SQUID. However, the integration of such a structure, which occupies a large area, hinders the down-scaling of the superconducting circuits.[7] Therefore, a truly gate-tunable Josephson junction (GTJJ) would be highly desirable for use in a variety of superconducting circuits.

In principle, the order parameter of a superconductor can be altered by applying a gate voltage[8] if its carrier concentration is low enough, like that of unconventional superconductors[9, 10]. Unfortunately, it has proved too challenging to develop a large-scale fabrication process for most unconventional superconductors when using common clean-room technologies[11]. Similarly, JJs based on the superconducting proximity effect can be also tuned using a local gate if the carrier concentration of the middle-layer material, e.g., semiconductor nanowires[12, 13], topological insulators[14], graphene[15-17], and other emerging materials[18-20], can be changed by applying an electric field. These structures have shown promise for applications in quantum circuits in the millikelvin regime[21, 22], but the critical current ($I_c$) and critical temperature of these proximity-effect-based JJs are too low for application to logic circuits. Another nanostructure called N-Tron[23] has been proposed for use as an interface between superconductor circuits and semiconductor circuits[24, 25] where a small current is injected through a third port to enable switching of a piece of superconducting nanowire[26]. Excitingly, recent experiments have shown that superconducting nanowires[27-29] and Dayem bridge junctions[30-33] made from conventional superconducting metals can also be tuned using an electric field in the sub-kelvin temperature regime while maintaining a constant superconducting order parameter. Apart from the unclear nature of the underlying mechanism[34-39], this phenomenon is not only interesting by itself but also is intriguing for superconducting circuit applications.

Here we report a GTJJ structure based on a niobium (Nb) 3D nano-bridge junction (NBJ)[40] with a voltage gate on top. The critical current of the NBJ can be tuned monotonically by applying a voltage to the gate. Our Nb-based GTJJ showed a working temperature of up to 6 K. The SQUID consisted of two NBJs in parallel and exhibited a magnetic flux modulation effect that is also tunable via the top gate voltage. The critical gate voltage was lowered by reducing the thickness of the insulation layer between the top gate and the NBJ. We believe that our GTJJ is potentially useful for building superconducting circuitry with high integration levels.

A schematic drawing of the structure of a GTJJ based on a 3D NBJ is shown in Figure 1a. The 3D NBJ consists of two thick niobium (Nb) banks connected via a thin nano-bridge. Both Nb banks are of ~100 nm in thickness and are separated by a $SiO_2$ insulating slit with a width of ~20 nm. The approximate deck dimensions of the nano-bridge comprise a thickness of 10 nm, length of 20 nm, and width of 50 nm. The 3D NBJ is covered by the voltage gate on the top and electrically isolated from the gate by a $SiO_2$ layer with a thickness of 130 nm. A false-colored scanning electron microscope (SEM) image of the GTJJ is shown in Figure 1b, where the top voltage gate is colored green. We characterized the junction by measuring current-voltage (*I-V*) curves with a constant voltage being applied to the gate. The sample preparation process and the measurement set-up are described in detail in the Methods section. The corresponding *I-V* curves at various values of $V_g$ are shown in Figure 1b. It is obvious that the critical current $I_c$ of the NBJ can be tuned monotonically to zero by increasing the gate voltage. The change in $I_c$ is symmetrical about the origin with reversal of the $V_g$ polarity. The normal resistance $R_n$ of the nano-bridge remained constant, which indicated that the physical properties of the nano-bridge remained unchanged. In Figure 1d, the full set of *I-V* curves obtained by sweeping $V_g$ was recorded as a contour plot. Here, $V_g$ was swept in both the forward and backward directions, and demonstrated the same behavior in both directions. The color scale shown in the figure represents the voltage drop across the NBJ. The superconducting and normal states of the NBJ are indicated by the blue and red colors, respectively. The boundary between the blue and red region is indicated by the solid line, which plots $I_c$ as a function of $V_g$. As shown, $I_c$ remained at a constant value when $V_g$ was low and formed a flat plateau in the $I_c$–$V_g$ curve. With increasing $V_g$, $I_c$ was then suppressed to zero gradually at the edge of the plateau. Here, we define the critical gate voltage $V_{gc}$ as the voltage when the $I_c$ of the NBJ drops to 90% of the corresponding value at $V_g = 0$ V.

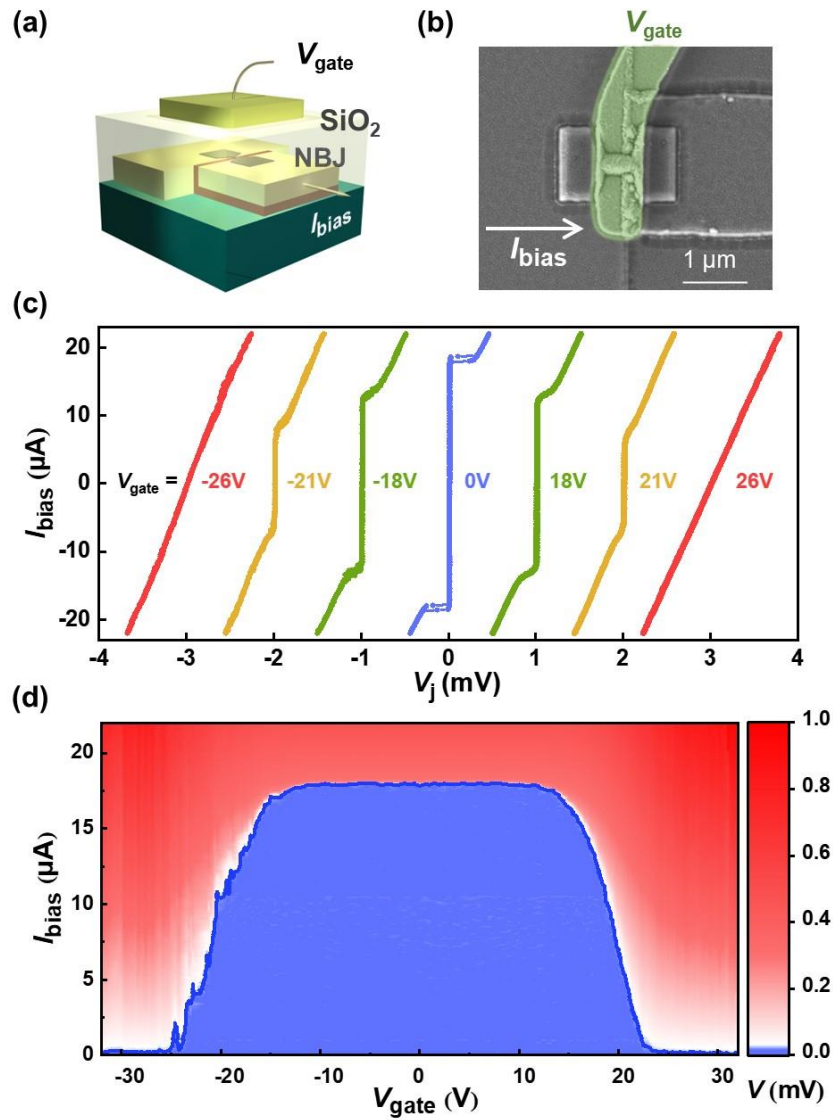

**Figure 1.** (a) Schematic drawing of the GTJJ structure. The top voltage gate is made from a Nb film and the transparent part represents the SiO$_2$ insulation layer that lies between the gate and the nano-bridge

junction. (b) False-colored scanning electron microscope (SEM) diagram of the device with the gate terminal colored in green. (c) Selected current-voltage curves measured at 4.2 K at various values of gate voltage $V_g$. For clarity, the curves were shifted 1 mV apart from each other along the $V_j$ axis. (d) Colored contour plot of the $V_j$ characteristic realized by sweeping the gate voltage $V_g$ and the bias current $I_b$. The color scale indicates the voltage drop $V_j$ across the NBJ. The blue solid line represents $I_c$ as a function of $V_g$, where $I_c$ is defined as the current when $V_j = 0.02$ mV.

In Figure 2a, we plotted $I_c$ as a function of $V_g$ with increasing temperature $T$. $V_g$ was swept in both the forward and backward directions for all curves, and no hysteresis was observed. At a constant $T$, the $I_c$ of the NBJ can be tuned effectively from a plateau to zero using $V_g$ in a similar manner to the characteristic at $T = 4.2$ K in Figure 1d. The $I_c$ plateau was lowered with increasing $T$ and reached zero at the critical temperature $T_c = 6$ K, as expected. However, the $I_c$ plateau broadened when $T$ increased from 4.2 K to 4.5 K. To observe this behavior clearly, $I_c$ was plotted as a function of $T$ at various $V_g$ in Figure 2b. We can see that $I_c$ dropped to zero at all $V_g$ at the same $T_c$ of 6 K. This behavior suggests that the order parameter of the NBJ remained constant with the changes in $V_g$. However, we can also see that $I_c$ initially spiked upward and then dropped off to zero with increasing $T$ when $V_g > V_{gc}$, while $I_c$ dropped monotonically with increasing $T$ when $V_g < V_{gc}$. In Figure 2c, we plotted $V_{gc}$ as a function of $T$. $V_{gc}$ also spiked upward initially and then decreased with increasing $T$. These results indicate that the superconductivity, which was diminished by the increase in temperature, was enhanced slightly by the application of the high $V_g$; this is similar to the result where $I_c$ increased at a high voltage when measured in a NbN nanowire[41]. The latter behavior was explained by a change in the vortex surface barrier on the nanowire under the influence of the electrical field. It is known that vortices will nucleate and pass through the NBJ, thus inducing voltage pulses, when a suitable bias current drives it from the superconducting state to the normal state. In our case, the increase in $T$ changes $\lambda/\xi$, where $\lambda/\xi$ is the ratio of the London penetration depth to the superconducting coherence length. At certain values of $\lambda/\xi$, the vortex surface barrier may require the application of a higher electrical field to tune the $I_c$ of the NBJ.

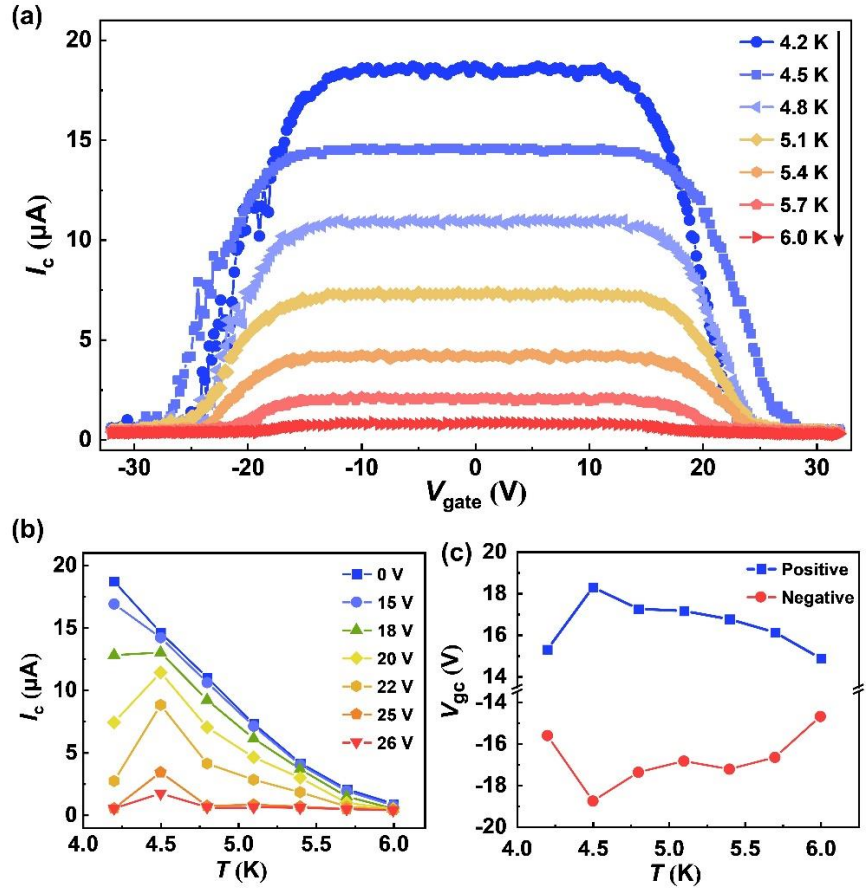

**Figure 2.** (a) Critical current $I_c$ as a function of gate voltage $V_g$ with increasing bath temperature $T$. (b) Dependence of $I_c$ on temperature $T$ at various $V_g$ values. $I_c$ spiked upward at 4.5 K when $V_g > V_{gc}$, where $V_{gc}$ is defined as the value of $V_g$ when $I_c$ drops to 90% of its value at 0 V. (c) $V_{gc}$ plotted as a function of $T$. $V_{gc}$ also spiked upward at 4.5 K.

To evaluate the influence of the leakage current $I_L$ between the gate and NBJ, we applied three voltage gates at different positions, as shown in Figure 3a. These gates remained 3 μm apart from each other, and the middle gate was placed on the top of the NBJ. We applied $V_g$ to all three gate terminals individually and then measured $I_L$ from each gate to the NBJ terminals. When $V_g = 30$ V was applied to the gate terminals from left to right, the $I_L$ values were 0.02 nA, 0.2 nA, and 0.2 nA, respectively (see the supporting information for the detail data). When compared with the current passing through the NBJ (~10 μA), each $I_L$ value was smaller by four to five orders of magnitude, and it would be impossible for such a current to influence the critical current significantly by itself. Furthermore, we plotted $I_c$ as a function of $V_g$ in Figure 3(b) for the case where $V_g$ was applied from the left to the right gate, as represented by the blue, red, and green colors, respectively. The only $I_c$–$V_g$ curve that shows a tunable $I_c$ was that when $V_g$ was applied to the middle gate. Neither application of $V_g$ to the left gate nor to the right gate altered the $I_c$ of the NBJ at all. The tunability of $I_c$ via $V_g$ disappeared when the voltage gate was dislocated from the top of the NBJ. Therefore, it is difficult to believe that either $I_L$ itself or the quasiparticle produced by it induced the suppression of $I_c$. In Figure 3c, we plotted both $V_{gc}$ and $V_g/I_L$ as a function of $d$, where $d$ is the vertical distance between the top gate and the NBJ. The values of $V_g/I_L$ for all $d$ were more than 50 GΩ and showed no apparent relationship with $d$. However, $V_{gc}$ decreased with decreasing $d$, which suggests that it is possible to reduce $V_{gc}$ further by lowering the $SiO_2$ layer thickness.[32]

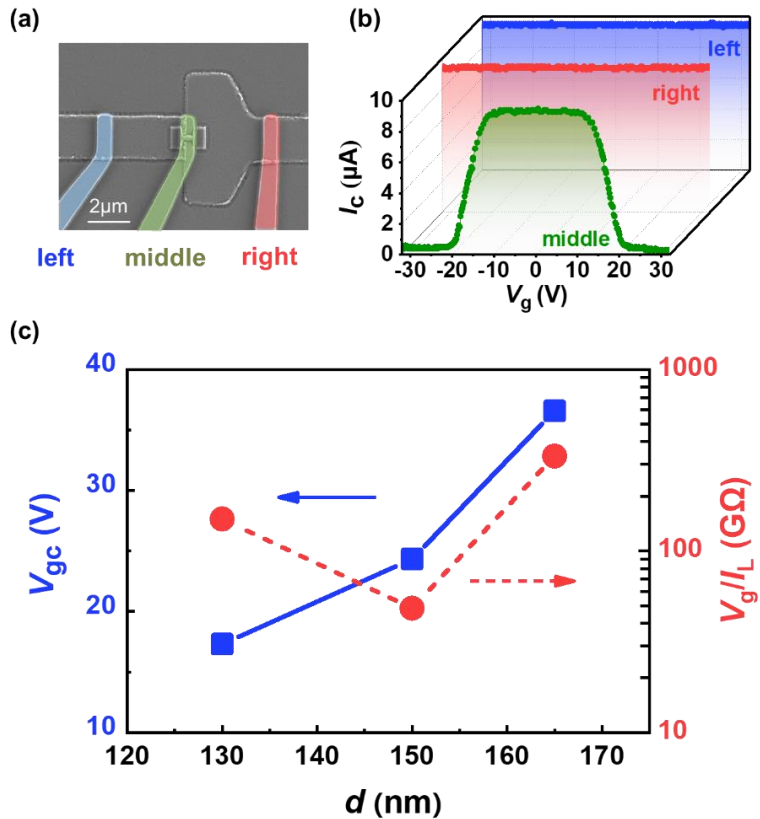

**Figure 3.** (a) False-colored SEM diagram of a device with gate terminals at different locations. The left, middle, and right gates were colored blue, green, and red, respectively. (b) Plots of $I_c$ as a function of $V_g$

when $V_g$ was applied to the left, middle, and right gate terminals individually. Suppression of $I_c$ was only observed when $V_g$ was applied to the middle terminal. (c) The blue squares plot the dependence of $V_{gc}$ on $d$, where $d$ is the thickness of the insulation layer between the voltage gate and the NBJ. The red circles plot $V_g/I_L$ at different $d$ and show no obvious dependencies.

To assess the Josephson effect of the NBJ under a constant $V_g$, we also arranged two NBJs in parallel to construct a SQUID with a voltage gate on top, as shown in Figure 4a. The SQUID loop is around $1\times1$ μm$^2$ in size. The dimensions of the NBJs are the same as those of the single NBJ tested previously. The $SiO_2$ layer thickness for this device was 165 nm. The magnetic flux that is coupled to the SQUID can be modulated using the current $I_{coil}$ in the coil line. As shown in Figure 4b, the entire flux modulation curve of the SQUID could be suppressed by increasing $V_g$. When $V_g = 0$ V, the SQUID showed a flux modulation depth FMD = $(I_{c\text{-max}} - I_{c\text{-min}})/I_{c\text{-max}}$ = 50%, where $I_{c\text{-max}}$ and $I_{c\text{-min}}$ are the critical currents of the SQUID at the points of constructive and destructive interference, respectively. A deep FMD indicated that the current phase relationship (CPR) of our NBJ does not deviate far from a sinusoidal function. When $V_g$ increased to 50 V, the modulation depth increased to 65%. It suggested that the CPR of NBJs evolved closer to a sinusoidal function because of the suppression of gate voltage.[42] Alternatively, the increase of FMD can also be attributed to the compensation of $I_c$ asymmetry through different voltage responses in the two parallel NBJs. In Figure 4c, we plotted all the $I_c$ values as a function of $I_{coil}$ by sweeping $V_g$ with $I_c$ along the z-axis. The values of $I_c$ are also indicated by the color bar. The $I_c$ values were periodic along the axis of $I_{coil}$ for all $V_g$ and were suppressed by increasing $V_g$, which confirms that the Josephson effect in the NBJs survived the gate-tuning process.

Theoretically, the electric field only penetrates for a few angstroms below the metal's surface because of the screening effect[43]. Additionally, the carrier concentration in the metallic NBJ is too high to allow the electric field to alter the order parameter. Here, we observed a constant $T_c$, which also suggested a constant order parameter. The underlying mechanism of this GTJJ has not been understood well to date.[29] There is a theory in which the mechanism is attributed to the superconducting Sauter-Schwinger effect[38], where an electrostatic field can generate two coherent excitations from a superconducting ground-state condensate that weakens the superconducting state. Recent reports have shown that the suppression effect may be related to the injection of high-energy electrons.[36, 39] It appears to be consistent with our results that showed the value of $V_{gc}$ dropping off with decreasing insulation layer thickness. However, $V_{gc}$ spiked upward at 4.5 K and then decreased with increasing temperature, which suggests that the electric field may also influence the surface vortex barrier of the NBJs and thus affect their critical currents. It may be intriguing to build a superconducting quantum transistor if one can achieve both suppression and enhancement of superconductivity by the gate voltage. Determination of whether the high-energy electrons break the Cooper pairs directly or the electric field influences the vortices across the NBJs would require a more microscopic investigation to be undertaken. Despite the obscure nature of its working mechanism, this GTJJ with its top gate structure has paved a way to fabricate a highly localized controller for all types of superconducting circuits for use in logic circuits or quantum computing.

In summary, we have developed a gate-tunable Josephson junction (GTJJ) based on a 3D Nb nano-bridge junction with a top voltage gate. Similar to other types of GTJJ, the junction's critical current can be tuned effectively using a third voltage gate. Furthermore, the proposed GTJJ shows an appreciable critical current with a relatively high $T_c$ of 6 K. The fixed $T_c$ with increasing $V_g$ indicates that the order parameter of the NBJ remains unchanged. The Josephson effect of the GTJJ is confirmed by the Josephson interference of a SQUID composed of two NBJs, which is also tunable via a gate voltage. With the top gate structure, it is possible to reduce $V_{gc}$ further by using a thinner insulation layer between the top gate and the NBJ. Therefore, we believe that this GTJJ structure will potentially be useful in the development of future superconducting circuits with high integration levels.

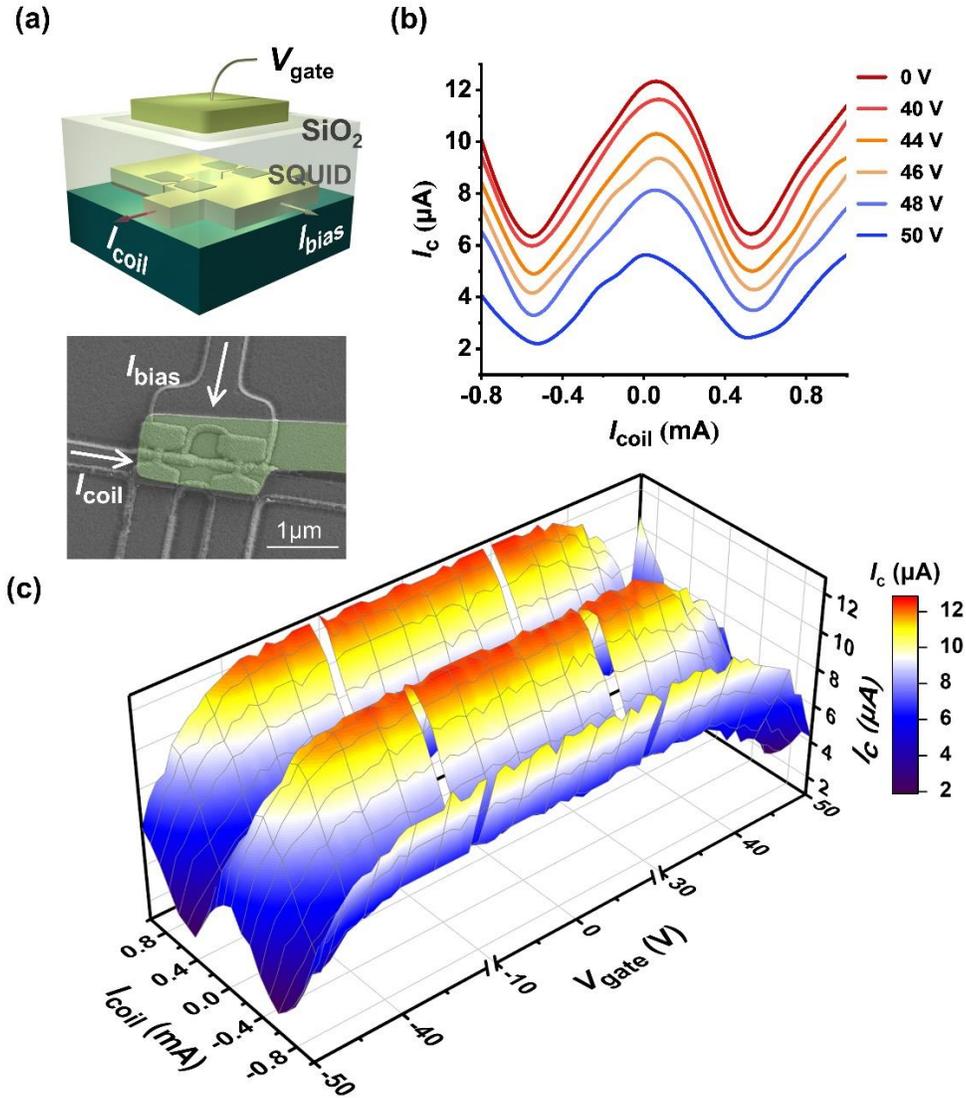

**Figure 4.** (a) False-colored SEM diagram and a 3D sketch of the SQUID with the top voltage gate. (b) Plotted magnetic flux modulation curves of the SQUID at various gate voltages. (c) Three-axis plot of the SQUID's $I_c$ as a function of $I_{coil}$ and $V_{gate}$. The $I_c$ values of the SQUID are also indicated by the color scale.

**Methods.**

*Sample Fabrication.* The fabrication process of the 3D Nb NBJ has been described previously in the literature[40]. In summary, the first 100-nm-thick Nb layer was deposited on a silicon wafer coated with 300 nm of $SiO_2$. After the first Nb layer was patterned by photolithography and reactive ion etching (RIE), we retained the photoresist layer on the wafer. Then, a 30-nm-thick $SiO_2$ layer and a second 110-nm-thick Nb layer was deposited and patterned again. By lifting off the $SiO_2$ layer and the second Nb layer from the first Nb layer, a vertical $SiO_2$ insulating slit with a width of 15 nm was formed between the two Nb banks. Next, we fabricated a 12-nm-thick Nb nano-bridge over the $SiO_2$ slit to connect the two thick Nb banks by electron-beam lithography. After the fabrication of the entire 3D NBJ structure, a thick $SiO_2$ layer and a Nb layer were deposited on top of the structure. The gate terminals were patterned out from the last Nb layer via photolithography and RIE.

*Measurement setup.* All tests were performed in liquid helium-filled Dewars using a customized 4 K probe. The *I-V* curves of the 3D NBJ and the SQUID devices were measured using the Keithley 2450 SourceMeter. The gate voltage $V_g$ was applied using a low-noise voltage source (SRS DC 205). The temperature variations were introduced by lifting the 4 K probe inside the liquid helium Dewar. The temperature sensor and the device were placed at the same horizontal level. In the SQUID modulation test, we used a current source (Keithley 2401) to generate the $I_{coil}$ required to drive the on-chip coil to generate the magnetic flux. The leakage current $I_L$ from the gate to the NBJ was measured using a low-noise current preamplifier (SRS SR570).

**Author Contributions**

SY and LC performed the experiments and collected the data. LC planned the research and wrote the paper under the supervision of ZW. YP, YW, DZ, GT, and XF participated in and assisted with the experiments. XL and LW assisted in the fabrication of devices under the supervision of WP. LZ and JR assisted in the data analysis. All authors approved the final version of the manuscript.**Acknowledgements**

The authors acknowledge support from the National Natural Science Foundation of China (Grant Nos. 62071458 and 11827805), the Strategic Priority Research program of the CAS (Grant No. XDA18000000), and the Young Investigator program of the CAS (Grant No. 2016217). The device fabrication was performed in the Superconducting Electronics Facility (SELF) of SIMIT.

**Additional Information**

*Competing Interests:* The authors declare no competing interests

# SUPPORTING INFORMATION

**1. The leakage current *vs* the gate voltage**

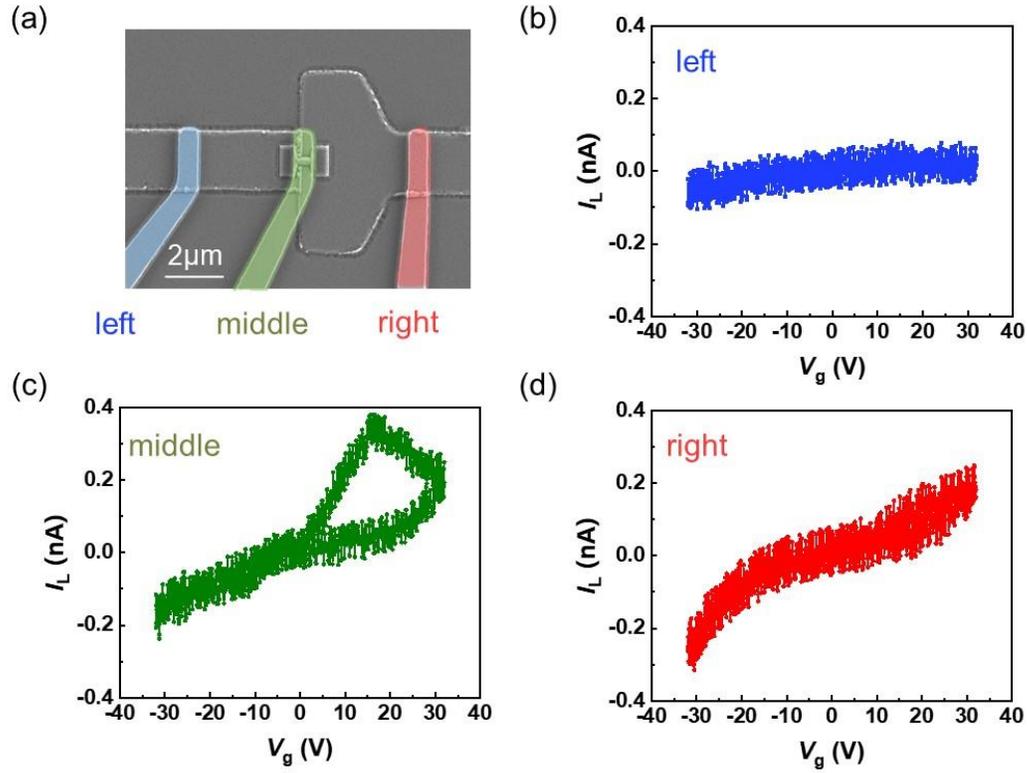

Figure S1 Plots of leakage current $I_L$ as a function of gate voltage $V_g$, where blue, red, and green curves correspond to the voltage applied on the left, middle, and right gate, respectively. The gate voltage $V_g$ was applied using a low-noise voltage source (SRS DC 205). The leakage current $I_L$ from the gate to the NBJ was measured using a low-noise current preamplifier (SRS SR570).